
\magnification1200
\font\BBig=cmr10 scaled\magstep2
\font\BBBig=cmr10 scaled\magstep3
\font\small=cmr7


\def\title{
{\bf\BBBig
\centerline{Spinor vortices}
\bigskip
\centerline{in}
\bigskip
\centerline{non-relativistic Chern-Simons theory}
}}

\def\foot#1{
\footnote{($^{\the\foo}$)}{#1}\advance\foo by 1
} 
\def\ccr{\cr\noalign{\medskip}}


\def\authors{
\centerline{
C. DUVAL\foot{D\'epartement de Physique, Universit\'e
d'Aix-Marseille II and Centre de Physique
\hfill\break
Th\'eorique, CNRS-Luminy, Case 907, F-13288 MARSEILLE,
Cedex 09 (France).
\hfill\break
e-mail:duval@cpt.univ-mrs.fr.}
P. A. HORV\'ATHY\foot{Laboratoire de Math\'ematiques et Applications,
Universit\'e de Tours, Parc de Grandmont,
F-37200 TOURS (France). e-mail: horvathy@balzac.univ-tours.fr}
L. PALLA\foot{Institute for Theoretical Physics,
E\"otv\"os University, H-1088 BUDAPEST,
\hfill\break
Puskin u. 5-7 (Hungary). e-mail: palla@ludens.elte.hu}}}

\def\runningauthors{Duval, Horv\'athy, Palla}

\def\runningtitle{Spinor vortices in Chern-Simons\dots}


\voffset = 1cm 
\baselineskip = 14pt 

\headline ={
\ifnum\pageno=1\hfill
\else\ifodd\pageno\hfil\tenit\runningtitle\hfil\tenrm\folio
\else\tenrm\folio\hfil\tenit\runningauthors\hfil
\fi
\fi}

\nopagenumbers
\footline={\hfil} 


\def\and{\qquad\hbox{and}\qquad}

\def\smallcirc{{\raise 0.5pt \hbox{$\scriptstyle\circ$}}} 
\def\smallover#1/#2{\hbox{$\textstyle{#1\over#2}$}} %
\def\2{{\smallover 1/2}}
\def\ccr{\cr\noalign{\medskip}}
\def\parag{\hfil\break} 
\def\={\!=\!}
\def\D{{D\mkern-2mu\llap{{\raise+0.5pt\hbox{\big/}}}\mkern+2mu}\ }

\def\L{{\cal L}}


\newcount\ch 
\newcount\eq 
\newcount\foo 
\newcount\ref 

\def\chapter#1{
\parag\eq = 1\advance\ch by 1{\bf\the\ch.\enskip#1}
}

\def\equation{
\leqno(\the\eq)\global\advance\eq by 1
}

\def\reference{
\parag [\number\ref]\ \advance\ref by 1
}

\ch = 1 
\eq = 1 
\foo = 1 
\ref = 1 


\title
\vskip 1.5cm
\authors
\vskip .25in

\parag
{\bf Abstract.}

{\sl
The non-relativistic `Dirac' equation of
L\'evy-Leblond is used to describe
a spin {\small 1/2} particle
interacting  with a Chern-Simons gauge field.
Static, purely magnetic, self-dual spinor vortices are
constructed.
The solution can be `exported' to a uniform magnetic background
field}.
\bigskip
%

\vfill\eject

The non-relativistic Chern-Simons
model of Jackiw and Pi [1] considers a
massive scalar field, $\Psi$, described
by the planar, gauged, non-linear Schr\"o\-din\-ger equation,
$$
iD_t\Psi=\left[
-{\vec{D}^2\over 2m}
-\Lambda\,\Psi^*\Psi\right]\Psi,
\equation
$$
while the electromagnetic field
and the current, $J^\alpha\=(\rho,\vec{J})$,
satisfy the field/current identity (FCI)
$$
\kappa B\equiv\epsilon^{ij}\partial_iA^j=
-e\varrho,
\qquad
\kappa E^i\equiv -\partial_iA^0-\partial_tA^i=
e\,\epsilon^{ij}J^j.
\equation
$$
The current and the fields are
coupled according to
$\varrho\=\Psi^*\Psi
$
and
$
\vec{J}\=\smallover1/{2im}[\Psi^*\vec{D}\Psi-\Psi(\vec{D}\Psi)^*],
$
where $D_\alpha=\partial_\alpha-ieA_\alpha$.
For the special value
$\Lambda=e^2/m\kappa$ of the non-linearity,
the static second-order equation (1) can be reduced
to the first-order 
equations
$
\big(D_1\pm iD_2\big)\Psi=0.
$
In a suitable gauge, this leads to Liouville's equation and can
therefore be solved explicitly [1]. Unlike the
vortices in the Abelian Higgs model [2], the
Jackiw-Pi vortices have non-zero electric as well as magnetic fields.

In this paper, we present new, {\it purely magnetic},
non-relativistic
solutions, constructed from a spin $\2$ (rather than scalar)
field with Chern-Simons coupling.
For this reason, we call them {\it spinor vortices}.
Being purely magnetic, our solutions differ from those
found by Leblanc et al. [3], who construct their solutions
by putting together
the JP solutions and their superpartners using supersymmetry.
Our vortices appear to be rather
the non-relativistic counterparts of the
relativistic fermionic vortices found by Cho et al. [4].

%
Following L\'evy-Leblond [5],
we describe non-relativistic, spin $\2$ fields by the $2+1$ dimensional
version of the non-relativistic `Dirac' equation
$$\left\{
\matrix{
(\vec{\sigma}\cdot\vec{D})\,\Phi
&+\hfill&2m\,\chi&=&0,
\ccr
D_t\,\Phi&+\hfill&i(\vec{\sigma}\cdot\vec{D})\,\chi\hfill&=&0,
\cr}\right.
\equation
$$
where  $\Phi$ and $\chi$ are
two-component `Pauli' spinors and
$(\vec{\sigma}\cdot\vec{D})=\sum\limits_{j=1}^2\sigma^jD_j$
with $\sigma^j$ denoting the Pauli matrices. These
spinors are coupled to the Chern-Simons gauge fields through
the mass (or particle) density  [5]:
$
\varrho=|\Phi|^2,
$
as well as through the spatial components of the current,
$
\vec{J}=i\big(\Phi^\dagger\vec{\sigma}\,\chi
-\chi^\dagger\vec{\sigma}\,\Phi\big),
$
so that the system (2)-(3) is self consistent.
Let us mention that
this coupled L\'evy-Leblond -- Chern-Simons system can be derived
from a $3+1$ dimensional {\it massless} Dirac -- Chern-Simons system
by a light-like dimensional reduction, in a way similar
to the one we used for a scalar field [6].
This
reduction \lq commutes' with the four-dimensional chirality operator
$$
\gamma^5=
\pmatrix{-i\sigma_3&0\cr0&i\sigma_3\cr},
\equation
$$
so that $\Phi$ and $\chi$ are {\it not} the chiral components
of the four-component spinor field $\psi=\pmatrix{\Phi\cr\chi\cr}$;
these latter are actually defined by
$
\2(1\pm i\gamma^5)\psi_\pm=\pm\psi_\pm.
$
Independently of this derivation it is easy to see that Eq. (3)
splits into two uncoupled systems for the two two-component spinor fields
$\psi_+$ and $\psi_-$.
Each of the chiral components separately describe (in
general different) physical phenomena in $2+1$ dimensions.
For the ease of presentation,
we keep, nevertheless, all four components of $\psi$.

Now the current can be written using (3) in the form:
$$
\vec{J}={1\over2im}\Big(
\Phi^\dagger\vec{D}\Phi-(\vec{D}\Phi)^\dagger\Phi\Big)
+\vec{\nabla}\times\Big({1\over2m}\,\Phi^\dagger\sigma_3\Phi\Big).
\equation
$$
Using
$(\vec{D}\cdot\vec{\sigma})^2=
\vec{D}^2+eB\sigma_3$,
we find that the component-spinors satisfy
$$\left\{
\matrix{
&iD_t\Phi&=&-{1\over2m}\Big[\vec{D}^2+eB\sigma_3\Big]\Phi,\hfill
\ccr
&iD_t\chi&=&-{1\over2m}\Big[\vec{D}^2+eB\sigma_3\Big]\chi
-\smallover{e}/{2m}\,(\vec{\sigma}\cdot\vec{E})\,\Phi.
\cr}\right.
\equation
$$
Thus, $\Phi$ solves a \lq Pauli equation', while
$\chi$ couples through the Darwin term,
$\vec{\sigma}\cdot\vec{E}$.
Expressing $\vec{E}$ and $B$ through the FCI, (2)
and inserting into our equations,
we get finally
$$\left\{
\matrix{
&iD_t\Phi=
&\Big[-{1\over2m}\,\vec{D}^2
+{e^2\over2m\kappa}\,|\Phi|^2\,\sigma_3
\Big]\Phi,\hfill
\ccr
&iD_t\chi=
&\Big[-{1\over2m}\,\vec{D}^2
+\smallover e^2/{2m\kappa}\,|\Phi|^2\,\sigma_3
\Big]\chi
-\smallover e^2/{2m\kappa}\,\big(
\vec{\sigma}\times\vec{J}\big)\Phi.
\cr}\right.
\equation
$$
If the chirality of $\psi$ is restricted to $+1$ (or $-1$),
this system describes
non-relativistic spin $+\2$ ($-\2$) fields interacting with a
Chern-Simons gauge field.
Leaving the chirality of $\psi$ unspecified, it
describes {\it two} spinor fields of spin $\pm\,\2$,
interacting with each other
and the Chern-Simons gauge field.

Since the lower component is simply
$
\chi=-(1/2m)(\vec{\sigma}\cdot\vec{D})\Phi,
$
it is enough to solve the
$\Phi$-equation.
For
$$
\Phi_+=\pmatrix{\Psi_+\cr0\hfill\cr}
\and
\Phi_-=\pmatrix{0\hfill\cr\Psi_-\cr}
\equation
$$
respectively --- which amounts to working with the $\pm$ chirality
components --- the \lq Pauli' equation for $\Phi$
reduces to
$$
iD_t\Psi_\pm=
\Big[-{\vec{D}^2\over2m}
\pm\lambda\,(\Psi_\pm^\dagger\Psi_\pm)\Big]\Psi_\pm,
\qquad
\lambda\equiv{e^2\over2m\kappa},
\equation
$$
which coincide with Eq. (1), but with non-linearities
$\pm\lambda$ --- the {\it half} of the
special value used by Jackiw and Pi.
For this reason, our solutions (presented below)
will be {\it purely magnetic}, ($A_t\equiv0$), unlike in the case
studied by Jackiw and Pi.

\goodbreak

In detail,
let us consider the static system
$$\left\{\eqalign{
&\Big[-{1\over2m}(\vec{D}^2+eB\sigma_3)-eA_t\Big]\Phi=0,
\ccr
&\vec{J}=-{\kappa\over e}\vec\nabla\times A_t,
\ccr
&\kappa B=-e\varrho,
\cr}\right.
\equation
$$
and try the first-order \lq self-dual' Ansatz
$$
\big(D_1\pm iD_2\big)\Phi=0.
\equation
$$
Eq. (11) makes it possible to replace $\vec{D}^2=D_1^2+D_2^2$
by $\mp eB$, then the first
equation in (10) can be written as
$$
\Big[-{1\over2m}eB(\mp 1+\sigma_3)-eA_t\Big]\Phi=0,
\equation
$$
while the current is
$$
\vec{J}
=
{1\over2m}\vec\nabla\times\Big[\Phi^\dagger(\mp1+\sigma_3)\Phi\Big].
\equation
$$

Now, due to the presence of $\sigma_3$,
Eq. (13) and the second equation in (10) can be solved
with a {\it zero} $A_t$ and $\vec{J}$:
by choosing
$\Phi\equiv\Phi_+$
($\Phi\equiv\Phi_-$) for the upper (lower) cases respectively
makes $(\mp 1+\sigma_3)\Phi$ vanish.
(It is readily seen from Eq. (12) that any solution has
a definite chirality).

The remaining task is to solve the SD conditions
$$
(D_1+iD_2)\Psi_+=0,\qquad{\rm or}\qquad(D_1-iD_2)\Psi_-=0,
\equation
$$
and $B\=-(e\varrho)/\kappa$, where $\varrho\=\vert\Psi_+\vert^2$
(or $\vert\Psi_-\vert^2$ respectively).
In the gauge
$\Psi_\pm\=\varrho^{1/2}$,
this yields [1]
$$
\vec{A}\=\pm{1\over2e}\vec{\nabla}\times\ln\varrho.
\equation
$$
Thus it reduces, in both cases, to the Liouville equation
$$
\bigtriangleup\ln\varrho=\pm{2e^2\over\kappa}\varrho.
\equation
$$
A normalizable solution is obtained for
$\Psi_+$ when $\kappa<0$, and for $\Psi_-$ when $\kappa>0$.
(These correspond precisely to having an attractive non-linearity
in Eq. (9)).
The lower components  vanish in both cases, as seen from
$\chi=-{1\over 2m}(\vec{\sigma}\cdot\vec{D})\Phi$.
Both solutions only involve
{\it one} of the $2+1$ dimensional spinor fields $\psi_\pm$, depending
on the sign of $\kappa$.

It is worth to point out that
inserting $A_t\=0$ into the first equation in (10)
yields the same equation as the one solved in  Ref. [1].
Remember, however, that
in the Jackiw~-~Pi construction, the FCI requires an electric
field such that
the $eA_t\Psi$ term coming from the covariant derivative
$D_t\Psi$ cancels {\it half} of the non-linearity.
Here we obtain the same effective theory with $A_t\=0$
and hence $\vec{E}\equiv 0$. Thus,
our solutions are \lq purely' magnetic.
Furthermore, since the total magnetic flux,
$
\int d^2x\,B\=
-(e/\kappa)\int d^2x\varrho
\equiv
-eN/\kappa,
$
is nonzero if $N\ne 0$, we  call our objects (spinor) vortices.

Eq. (16) can be analyzed following Ref. [1].
The general solution of the Liouville equation is
$
\varrho=\mp(4\kappa/e^2)|f'(z)|^2(1+|f(z)|^2)^{-2}
$
with $f(z)$ complex analytic.
A radially
symmetric solution is provided, for example, [1] by:
$$
\varrho_n(r)=\mp{4n^2\kappa\over e^2r^2}\left(({r_0\over r})^n
+
({r\over r_0})^n\right)^{-2},\qquad z=r{\rm e}^{i\theta},
\equation
$$
where $r_0$ and $n$ are two free parameters.
(The single valuedness of $\Psi_\pm$  requires $n$
to be an integer, though [1]).
Integrating $\varrho_n$ over all two-space yields
$
N=4\pi n{\vert\kappa\vert}/e^2.
$

The physical properties as symmetries and conserved quantities
can be studied by noting that our
equations are in fact obtained by variation of
the $2+1$-dimensional action $\int\!d^3x\L$ with
$$\eqalign{
&{\cal L}={\kappa\over4}\epsilon^{\mu\nu\rho}A_\mu F_{\nu\rho}+
\ccr
&\Im\left\{\bigl[
\psi_+^\dagger
\big(\Sigma^t_+D_t+\Sigma^i_+D_i-2im\Sigma^m_+\big)\psi_+
\bigr]
+\bigl[
\psi_-^\dagger
\big(\Sigma^t_-D_t+\Sigma^i_-D_i-2im\Sigma^m_-\big)\psi_-
\bigr]\right\}
\cr}
\equation
$$
where, with some abuse of notation, we identified the chiral
components $\psi_\pm$ with 2-component spinors and introduced
the $2\times 2$ matrices
$$\matrix{
\Sigma^t_+=\Sigma^t_-=\hfill
&\2(1+\sigma_3),\qquad\hfill
&\Sigma^m_+=\Sigma^m_-=\hfill
&\2(1-\sigma_3),\hfill
\ccr
\Sigma^1_+=-\sigma_2,\hfill
&\Sigma^2_+=\sigma_1,\hfill
&\Sigma^1_-=-\sigma_2,\hfill
&\Sigma^2_-=-\sigma_1.\hfill
\cr}
\equation
$$
Note that the matter action decouples into chiral components,
consistently with the decoupling of
the L\'evy-Leblond equation (3).

The action (18) can be used to
show that the coupled L\'evy-Leblond --- Chern-Simons system is,
just like its scalar counterpart, Schr\"odinger symmetric [1],
proving that our theory is indeed non-relativistic.
A conserved energy-momentum tensor can be constructed and used to
derive conserved quantities. One finds that
the `particle number' $N$ determines the actual values of all
the conserved charges: for (17), e.g.,
the magnetic flux, $-eN/\kappa$,
and the mass, ${\cal M}=mN$,
are the same as for the scalar soliton of [1].
The total angular
momentum, however, can be shown to be $I=\mp N/2$,
{\it half} of the corresponding value
for the scalar soliton.
As a consequence of self-duality,
our solutions have {\it vanishing energy}, just like the ones of
Ref. [1].

Our vortices are hence similar to the
non-relativistic scalar solitons
of Jackiw and Pi. On the other hand,
the similar flux/angular momentum
ratio and the vanishing $A_t$ indicate that they are just as akin
to the relativistic fermionic vortices found by Cho et al. [4].

\goodbreak

Physical applications as the fractional quantum Hall effect
[7] would require to extend the theory to background fields.
Solutions in
a background uniform magnetic or harmonic force field can be
obtained by `exporting' the empty-space solution,
just like for scalars [8].
Let ${\cal A}_t\equiv-U$ and $\vec{\cal A}$ denote the potentials of
an external
electromagnetic field and
consider the modified L\'evy-Leblond equation
$$\left\{
\matrix{
(\vec{\sigma}\cdot\vec{D}^{ext})\,\Phi^{ext}
&&&+\hfill&2m\,\chi^{ext}&=&0,
\ccr
D^{ext}_t\,\Phi^{ext}&+&{ie\over4m}{\cal B}\sigma_3\Phi^{ext}
&+\hfill&i(\vec{\sigma}\cdot\vec{D}^{ext})\,\chi^{ext}\hfill&=&0,
\cr}\right.
\equation
$$
where
${\cal B}=\vec{\nabla}\times\vec{\cal A}
$
is the external magnetic field and the new covariant derivative is
$
D^{ext}_\alpha=\partial_\alpha-ieA_\alpha-ie{\cal A}_\alpha.
$
Note that we have also included the anomalous term
${ie\over4m}{\cal B}\sigma_3\Phi^{ext}$, which is the reduction of
the  anomalous term
$\smallover1/{16}{\cal F}_{ij}[\gamma^i,\gamma^j]\gamma^t\psi$ arising
in $3+1$ dimensions [6].
Then the `upper' component  $\Phi^{ext}$ solves the
`Pauli' equation with anomalous magnetic moment
$$
iD^{ext}_t\,\Phi^{ext}=
-{1\over2m}\Big[\big(\vec{D}^{ext}\big)^2+eB\sigma_3\Big]\Phi^{ext}
-{e{\cal B}\over4m}\sigma_3\Phi^{ext}.
\equation
$$

In a constant ${\cal B}$-field,
for example, set ${\cal A}_i=-\2\epsilon_{ij}{\cal B}x^j$, and
$$\eqalign{
&\psi^{ext}(\vec{x}, t)=\hfill\ccr
&\;
{e^{-(im\omega r^2\tan\omega t)/2}\over\cos\omega t}\,
\pmatrix{
e^{i\omega t\sigma_3/2}\hfill
&0
\ccr
i{\omega\over2}\big[\tan\omega t
(\vec{\sigma}\cdot\vec{x})
-(\vec{\sigma}\times\vec{x})\big]e^{i\omega t\sigma_3/2}
\quad\hfill
&{e^{i\omega t\sigma_3/2}\over\cos\omega t}\hfill
\cr}
\psi^0(\vec{X},T),
\cr\cr
&A^{ext}_\alpha=\partial_\alpha X^\beta A^0_\beta,
\qquad
\cr}
\equation
$$
with
$$
\vec{X}=[\cos\omega t]^{-1}R^{-1}(\omega t)\,\vec{x},
\qquad
T=\omega^{-1}{\tan\omega} t.
\equation
$$
Here $\omega=e{\cal B}/2m$ and $R(\theta)$
is the matrix of a rotation by angle $\theta$ in the plane:
$\vec\sigma\cdot(R\vec{x})=a(\vec\sigma\cdot\vec{x})a^{-1}$ with
$a(\theta)\equiv\exp(i\theta\sigma_3/2)$.
Then a straightforward calculation shows that (22) solves the
external-field equation (20) whenever
$(\psi^0, A_\alpha^0)$ is a solution in `empty' space
i.e. with no external field.
Note that while the `upper' component $\Phi$ transforms with the same
`time-dependent dilation' factor $[\cos\omega t]^{-1}$
as a scalar field,
the `lower' component, $\chi$,
has a conformal factor $[\cos\omega t]^{-2}$
as well as an inhomogenous part, which is linear in $\Phi$.
The time derivative of the rotation
matrix in Eq. (22) compensates in particular the anomalous term
in (21), while in Eq. (20) for this compensation
the inhomogenous part of the lower component
is also needed.
Applying the transformation (22) to the static spinor
vortices Eqs. (17,15) yields in particular
time-dependent background-field
solutions with non-vanishing `lower' component
as well as a (Chern Simons) electric field.

\goodbreak

Let us mention in conclusion, that the derivation of the coupled
L\'evy-Leblond -- Chern-Simons system based on
light-like dimensional reduction makes
also transparent its Schr\"odinger symmetry. In this procedure, the
four-dimensional chirality operator, Eq. (4), and
the anomalous term in the external-field, (20), arise naturally.
Finally, the external-field equation (20)
and the `solution-exporting' formula (22) both
have a nice geometric meaning.
These points are explained in Ref. [6].

\vskip4mm
\centerline{\bf\BBig References}

\reference
R. Jackiw and S-Y. Pi,
Phys. Rev. Lett. {\bf 64}, 2969 (1990);
Phys. Rev. {\bf 42}, 3500 (1990);
For a review, see Prog. Theor. Phys. Suppl. {\bf 107}, 1 (1992).

\reference
H. B. Nielsen and P. Olesen, Nucl. Phys. {\bf B61}, 45 (1973)

\reference
M. Leblanc, G. Lozano and H. Min,
Ann. Phys. (N.Y.) {\bf 219}, 328 (1992).

\reference
Y. M. Cho, J. W. Kim, and D. H. Park,
Phys. Rev. {\bf D45}, 3802 (1992).

\reference
J-M. L\'evy-Leblond, Comm. Math. Phys. {\bf 6}, 286 (1967).

\reference
C. Duval, P. A. Horv\'athy and L. Palla,
{\it Spinors in non-relativistic Chern-Simons theory},
(in preparation);
%
Phys. Lett. {\bf B325}, 39 (1994);
Phys. Rev. {\bf D50}, 6658 (1994).

\reference
S. M. Girvin, {\it The quantum Hall Effect},
ed. R. E. Prange and S. M. Girvin. Chap. 10.
(Springer Verlag, New York (1986);
S. M. Girvin and A. H. MacDonald,
Phys. Rev. Lett. {\bf 58}, 1252 (1987);
S. C. Zhang, T. H. Hansson and S. Kivelson,
Phys. Rev. Lett. {\bf 62}, 82 (1989).

\reference
U. Niederer, Helv. Phys. Acta {\bf 46}, 192 (1973);
Z. F. Ezawa, M. Hotta and A. Iwazaki,
Phys. Rev. Lett. {\bf 67}, 411 (1991);
R. Jackiw and S-Y. Pi,
Phys. Rev. Lett. {\bf 67}, 415 (1991);
Phys. Rev. {\bf D44}, 2524 (1991).

\bye